\documentclass[aps,prd,epsf,showpacs,amsmath,twocolumn,amssymb,
graphics,10pt]{revtex4}
\usepackage{subfigure}
\usepackage{graphicx}
\usepackage{dcolumn}
\usepackage{bm}

\newcommand{\be}{\begin{equation}}
\newcommand{\ee}{\end{equation}}
\newcommand{\ba}{\begin{eqnarray}}
\newcommand{\ea}{\end{eqnarray}}
\newcommand{\ban}{\begin{eqnarray*}}
\newcommand{\ean}{\end{eqnarray*}}

\newcommand{\eq}[1]{(\ref{#1})}

\begin{document}
\title{Finite escape fraction for ultrahigh energy collisions 
around Kerr naked singularity }
\author{Mandar Patil \footnote{ Electronic address: mandarp@tifr.res.in}
and  Pankaj S. Joshi \footnote{ Electronic address: psj@tifr.res.in}}
\affiliation{Tata Institute of Fundamental Research\\
Homi Bhabha Road, Mumbai 400005, India}

\begin{abstract}
We investigate here the issue of observability of high energy 
collisions around Kerr naked singularity and show that the results 
are in contrast with the Kerr black hole case. We had shown earlier
that it would be possible to have ultrahigh energy collisions 
between the particles close to the location $r=M$ around the Kerr 
naked singularity if the Kerr spin parameter transcends the unity 
by an infinitesimally small amount $a\rightarrow 1^+$. The collision 
is between initially ingoing particle that turns back as an outgoing 
particle due to angular momentum barrier, with another ingoing
particle. We assume that two massless particles are produced in 
such a collision and their angular distribution is isotropic in the 
center of mass frame. We calculate the escape fraction for the massless 
particles to reach infinity. We show that escape fraction is  
finite and approximately equal to half for ultrahigh energy collisions. 
Therefore the particles produced in high energy collisions would 
escape to infinity providing signature of the nature of basic
interactions at those energies. This result is in contract with the 
extremal Kerr black hole case where almost all particles produced 
in high energy collisions are absorbed by the black hole, thus 
rendering the collisions unobservable.

\end{abstract}
\maketitle

\section{Introduction}
Banados, Silk and West demonstrated an intriguing possibility 
of collisions of particles with arbitrarily large 
center of mass energy around extremal Kerr black hole \cite{BSW1}.
The collisions were between two ingoing particles that start from 
rest at infinity and fall freely 
towards the black hole. The collision takes place at a location 
that is close to the event horizon.
It turns out that in order for collision to be energetic the 
angular momentum of one of the particles 
must be finetuned to a specific value \cite{Berti}. This particle 
approaches the event horizon asymptotically. 
Thus the proper time required for the 
collision is infinite. A similar process was also studied in the 
context of many other known examples of black holes
by different authors. While analyzing the possible implications 
of the high energy collision process for the distant observers at infinity
the escape fraction was subsequently calculated. Escape fraction tells 
us the fraction of particles created in the 
high energy collisions that manage to escape to infinity. It turns 
out that since the 
collision takes place at a location close to the event horizon
most of the particles produced are absorbed by the black hole. Thus 
the escape fraction in the case of a black hole turns out to be 
vanishingly small \cite{BSW2,nn}. Thus although ultrahigh energy 
collisions would take place around extremal Kerr black holes, they 
would be inconsequential from the perspective of observational 
implications and in such a case the physics at high energies 
would remain unprobed.

We studied earlier the process of high energy collisions in 
the context of Kerr naked singularity \cite{Patil}. We showed that 
it would be 
possible to have ultrahigh energy particle collisions around Kerr 
naked singularity provided the Kerr spin parameter is
larger than but arbitrarily close to unity. Collisions take place at 
a location which is away from the singularity.
The collision is between ingoing and outgoing particles. Outgoing 
particles naturally arise as initially ingoing particles 
are reflected back due to the potential barrier offered by the angular 
momentum of the particle in the equatorial 
plane and due to the repulsive nature of gravity along the axis 
of symmetry. We also studied the similar process 
in the context of Reissner-Nordstr\"om naked singularity when 
charge parameter 
transcends the mass parameter by infinitesimal amount \cite{Patil2}. 
In this case, exploiting the spherical symmetry we could replace
the particles by thin spherical shells and take into account the
effect of self-gravity of colliding particles 
on the center of mass energy of collisions. We showed that in a 
reasonable astrophysical scenario the upper
bound on the center of mass energy of collision between two 
proton like particles was much larger than the Planck scale.
Whereas a similar calculation in the context of black holes showed 
the upper bound on center of mass energy 
of collision to be much smaller then the Planck energy. We also 
showed that neither black holes 
nor naked singularities are necessary for high energy collisions to occur.
We demonstrated this point in the context of the Bardeen spacetime. 
It turns out that ultrahigh energy collisions can take place
in the spacetimes that do not contain black hole \cite{Patil3}.

In this paper we address the issue of observability of high energy 
collisions in the absence of an event horizon. We 
specifically work in the context of Kerr naked singular spacetime. 
We calculate the escape fraction for the 
particles produced in the high energy collisions around Kerr naked 
singularity. Following \cite{BSW2} we make various assumptions.
The collision considered is between two identical massive particles 
starting from rest at infinity. The collision, 
as stated before, is between ingoing and outgoing particle. We assume 
that the particles move on the equatorial plane.
Collision products are taken to be two massless particles. In the 
center of mass frame colliding particles move in 
in the opposite directions with same speed. In this frame two massless 
particles produced move in the opposite directions 
with equal magnitude of momenta. We assume that the two massless particles 
also travel in the equatorial plane. The distribution of the direction 
of motion of massless particles with respect to the direction of 
colliding particles is taken to isotropic. We focus 
attention on only one of the particles as we intend to compute the 
order of magnitude of escape fraction.

We show that the escape fraction for the collision products to 
escape will be finite and would be approximately half.
This implies that around half of the particles produced eventually hit the 
singularity and disappear while remaining half
manage to escape away to infinity. Thus significant fraction of collision 
products escape to infinity and would carry the information about 
nature of basic interactions at extremely large energy scales. This 
is in contrast with the black hole case where high 
energy collisions that take place in the vicinity of event horizon 
would be unobservable.

\section{Kerr geometry}
In this section we briefly describe the Kerr naked singularity geometry.
We work in the units where the gravitational
radius, speed of light and mass of the colliding particles is 
set to unity $r_{g}=\frac{GM}{c^2}=c=\mu=1$.

The Kerr metric when restricted to the equatorial plane, in Boyer-Lindquist
coordinates is given by,
\begin{equation}
\begin{split}
 ds^2=&- \left( 1- \frac{2}{r}\right)dt^2- \frac{4ra}{\Sigma}
dt d\phi 
+\left(\frac{r^2}{\Delta}\right)dr^2+ r^2 d\theta^2  \\  &+
\left(r^2+a^2+\frac{2a^2}{r}\right) d\phi^2 \label{KBL1}
\end{split}
\end{equation}
where $\Delta= r^2+a^2-2r$ and $a$ is the angular momentum parameter.
When $a>1$, event horizon is absent in the spacetime. However, 
there is a singularity located 
at $r=0, \theta=\frac{\pi}{2}$, which turns out to be visible for 
the asymptotic observer. Thus
the Kerr solution represents a naked singularity.

We now analyze the geodesic motion in the spacetime. 
Let $U=\left(U^t,U^r,U^{\theta},U^{\phi}\right)$ be the velocity 
of the particle. We assume for simplicity that particle follows geodesic 
motion in the equatorial plane $\theta=\frac{\pi}{2}$ so that $U^{\theta}=0$.
The Kerr spacetime admits Killing vectors $\partial_t,\partial_{\phi}$ 
and there 
are corresponding conserved quantities $E=-U.\partial_{t}$, $L=U.\partial_{\phi}$
associated with the geodesic motion which are interpreted as the 
conserved energy and orbital
angular momentum per unit mass. Using these relations, 
$U^t,U^{\phi}$ can be written down and 
$U^r$ can be obtained using the normalization condition for 
velocity $U.U=-m^2$. Here $m=\mu=1$
for the massive particle and $m=0$ for the massless particle. 
The velocity components are given by,
\begin{eqnarray}
\nonumber
U^{t}=\frac{1}{\Delta}\left[\left(r^2+a^2+\frac{2 a^2}{r}\right)E
-\frac{2a}{r} L \right] \\
U^{\phi}= \frac{1}{\Delta} \left[\left(1-\frac{2}{r}\right)L
+\frac{2a}{r}E\right], \; U^{\theta}=0 \label{Ur}
\end{eqnarray}
\begin{equation}
U^{r}=u\sqrt{E^2-m^2+\frac{2m^2}{r}-\frac{\left(L^2-a^2
\left(E^2-m^2\right)\right)}{r^2}
+\frac{2\left(L-aE\right)^2}{r^3}}
\nonumber
\end{equation}
where $u=\pm 1$ stands for radially outgoing and
ingoing particles respectively. We shall deal with the colliding particles 
that start from rest at infinity. For such particles it turns out that $E=1$.

\section{Ultrahigh energy collisions}
\begin{figure}
\begin{center}
\includegraphics[width=0.4\textwidth]{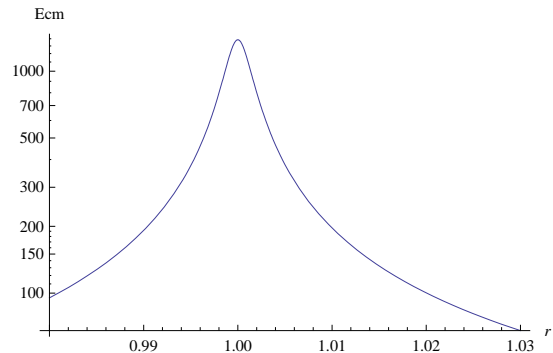}
\caption{\label{fg1}
The variation of center of mass energy
of collision between the ingoing and outgoing particles
with angular momenta $L_{1}=0.4,L_{2}=1.4$ with radius around $r=1$
in a spacetime containing naked singularity with spin
parameter close to unity, with $a-1=\epsilon=10^{-6}$.
}
\end{center}
\end{figure}

In this section we briefly describe the process of ultrahigh 
energy particle collisions 
in the Kerr naked singularity geometry with the spin parameter 
transcending unity by an extremely small 
amount $a-1=\epsilon=1^+$ \cite{Patil}.

We consider collision between ingoing and outgoing particles at $r=1$.
We assume that both the particles start from infinity at rest. 
Thus the conserved energy
per unit mass for each of the particles is $E=1$. One of the initially
ingoing particles must turn back as an outgoing particle at the 
radial location $r<1$.
The angular momentum per unit mass of this particle must lie in the range 
$2\left(-1+\sqrt{1+a}\right)\leq L < \left(2-\sqrt{2a^2-2}\right)$.
The lower limit on its value corresponds to the angular momentum it must have
for it to turn back at all, whereas the upper limit comes from the 
restriction that 
it must turn back at the radial location $r<1$. The second particle is 
an ingoing particle
at the collision location. The requirement that it should not turn
back before it reaches $r=1$ puts
the following restriction on its angular momentum 
$L < \left(2-\sqrt{2a^2-2}\right)$ \cite{Patil}.

The center of mass energy of collision between the
particles with velocities $U^1,U^2$ is given by
\begin{eqnarray}
 E_{c.m.}^2=2m^2\left(1-g_{\mu\nu}U_{1}^{\mu}U_{2}^{\nu}\right)
\label{Ecm}
\end{eqnarray}
Approximate expression for the center of mass energy of collision
in the situation that we described is given by 
\begin{eqnarray}
 E_{c.m.}^2 \simeq \frac{2T_1T_2}{\epsilon}
\end{eqnarray}
where $T_1,T_2$ are factors that can be written down in terms 
of the energy and angular momenta
of the particles and they are of $O(1)$. Clearly the center 
of mass energy of collision is large
in the limit $\epsilon=a-1 \rightarrow 0$.
The variation of center of mass energy in the vicinity
of $r=1$ is shown in Fig 1 for Kerr geometry 
with $\epsilon=10^{-6}$.

\section{Escape fraction for collision products}
We now compute the escape fraction for the collision products. 
We assume that two 
identical massive particles each with mass $\mu(=1)$ 
participate in a collision and 
produce two massless particles. In the center of mass frame 
colliding particles travel in the 
opposite direction with equal magnitude of three-momenta. This 
is also the case with massless particles produced in the collision.
The distribution of massless particles in the center of mass 
frame is assumed to be isotropic for simplicity and we focus
only on one of the particles since we intend to calculate the order 
of magnitude of escape fraction. 
We assume that the massless particles travel in the equatorial plane. 
We first calculate the conditions under which massless particle 
will be able to escape to infinity. 
We then move over to the center of mass frame. The conditions 
for particles to escape can be translated 
to the escape cones in the center of mass frame within 
which particles must travel if it were to escape to infinity.
Sum of angles of these cones divided by $2\pi$ will be the 
estimate of the escape fraction.

\subsection{Escape conditions for massless particles}
We now derive the conditions for massless particles to escape 
to infinity. Null geodesics
are characterized by the ratio of conserved energy and angular 
momentum $\frac{L}{E}$, rather than by both conserved 
energy and angular momentum as in the case of timelike geodesics. 
This can be seen from \eq{Ur} by dividing it 
throughout by $E$ and choosing a new affine parameter 
$\tilde{\lambda}=E\lambda$ where $\lambda$ is the old affine parameter. 
Equivalently, one can choose $E=1$.

The angular momentum required for the particle to turn back at 
the radial coordinate $r$
can be obtained from \eq{Ur} and is given by
\begin{equation}
\begin{split}
 b(r)&=L_{1}(r)=\frac{1}{r-2}\left(-2a+\sqrt{r^4+a^2r^2-2r^3}\right) \\
     &=L_{2}(r)=\frac{1}{r-2}\left(-2a-\sqrt{4r^4+a^2r^2-2r^3}\right)
\end{split}
\end{equation}
Two branches of $b(r)$ namely $L_{1}(r),L_{2}(r)$ are plotted 
in Fig.2. The lower branch of
$L_{2}(r)$ admits a maximum $L_{2max} \simeq -7$ around $r=r_{max}\simeq
4$ for Kerr spin parameter close to unity. The massless particle
produced in a collision at $r$ could either be ingoing or
outgoing corresponding to $u=\pm 1$. The particle will escape
to infinity if
\begin{eqnarray}
\label{escape}
r<r_{max},u=+1, L_{2,max}<L<L_{1}(r)\\
\nonumber
r<r_{max},u=-1, a<L<L_{1}(r)\\
\nonumber
r>r_{max},u=+1, L_{2}<L<L_{1}(r)\\
\nonumber
r>r_{max},u=-1, a<L<L_{1}(0)\\
\nonumber
r>r_{max},u=-1, L_{2}(r)<L<L_{2,max}
\end{eqnarray}
The first condition in \eq{escape} indicates that if the collision 
takes place at radial location $r$ and if the massless
particle produced in the collision is to be radially ingoing then 
it would escape to infinity if its angular momentum lies in the 
range $L_{2,max}<L<L_{1}(r)$.
This is evident from Fig.2. Other four conditions can be interpreted 
similarly. 
These conditions can be translated into the escape fraction as we 
demonstrate below.
\begin{figure}
\begin{center}
\includegraphics[width=0.4\textwidth]{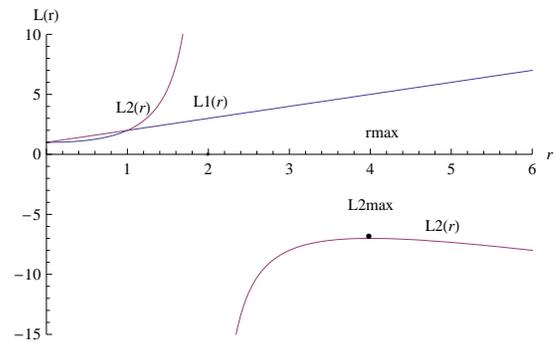}
\caption{\label{fg2}
The variation of angular momentum $b(r)=L_{1}(r),L_{2}(r)$
required for the massless particle to turn back from radius
$r$. The lower branch of $L_{2}(r)$ admits a maximum
$L_{2}=L_{2max}\approx -7$ around $r=r_{max}\approx 4$ for Kerr
spin parameter $a\approx 1$.  }
\end{center}
\end{figure}

\subsection{Center of mass frame}
We now consider the procedure to move over to the center of mass frame.
We first introduce locally non-rotating frame \cite{Bardeen}. 
Given a vector in Boyer-Lindquist 
coordinate system $V^{\mu}$, its components in the locally 
non-rotating frame $V^{\tilde{\mu}}$ are given by
\begin{equation}
V^{\tilde{\mu}}=e^{~\tilde{\mu}}_{\nu} V^{\nu}
\end{equation}
where
\begin{equation}
\nonumber
e^{~\tilde{\mu}}_{\nu}=\begin{pmatrix}
\sqrt{\frac{g_{t\phi}^2-g_{tt}g_{\phi\phi}}{g_{\phi\phi}}}&0&0&\frac
{g_{t\phi}}{\sqrt{g_{\phi\phi}}} \\
0&\sqrt{g_{rr}}&0&0 \\
0&0&\sqrt{g_{\theta \theta}}&0 \\
0&0&0&\sqrt{g_{\phi\phi}}
\end{pmatrix}
\end{equation}
in the above the $g_{\mu\nu}$ being the Kerr metric.
We then make a rotation in the $\tilde{r}-\tilde{\phi}$ plane 
in the locally non-rotating frame so that the net 
three velocity of two colliding particles is oriented along the 
new radial direction.
We further make a boost along new radial direction so that 
all the components of total three velocity now vanish.
This is the center of mass frame.

The transformation to center of mass frame from
locally nonrotating frame can be given as
\begin{equation}
 V^{\hat{\mu}} =\Lambda^{\hat{\mu}}_{~\tilde{\nu}}V^{\tilde{\nu}}
\end{equation}
where the transformation matrix is composed of rotation and boost
\begin{equation}
\Lambda=\Lambda_{boost}\Lambda_{rot}=\begin{pmatrix}
\gamma & -\beta\gamma \cos{\alpha} & 0 & -\beta\gamma \sin{\alpha}  \\
-\beta \gamma & \gamma \cos{\alpha} & 0 & \gamma \sin{\alpha}  \\
0 &0&1&0 \\
0 & -\sin{\alpha} & 0& \cos{\alpha} \\
\end{pmatrix}
\end{equation}

The boost parameter $\beta$ and rotation parameter 
$\alpha$ are given in terms of components of total velocity 
$U_{tot}=U_1+U_2$ in the locally non-rotating frame as 
\begin{equation}
\begin{split}
 \beta=&\left(\frac{\sqrt{U^{\tilde{r}2}_{tot}+U^{\tilde{\phi}2}_{tot}}}
{U^{\tilde{t}}_{tot}}\right) \\
\alpha=&\arccos{\left(\frac{U^{\tilde{r}}_{tot}}
{\sqrt{U^{\tilde{r}2}_{tot}+U^{\tilde{\phi}2}}}\right)}
\end{split}
\end{equation}

Given the energy and angular momentum parameters of the 
colliding particles one can write down the velocity
components in the Boyer-Lindquist coordinate system and also in 
the locally non-rotating orthogonal tetrad.
The rotation and boost parameters can be written down once 
components of the colliding particles are known 
in the non-rotating frame. Thus we can write down transformation 
laws that will allow us to write down components 
of any vector in the center of mass frame given its components 
in the Boyer-Lindquist coordinate system.

\subsection{Escape Fraction}
\begin{figure*}[htb]
    \centering
    \hspace{-2.5cm}
    \begin{minipage}[h]{6cm}
      \centering
      \subfigure[]
      {
        \resizebox{8cm}{!}{\includegraphics{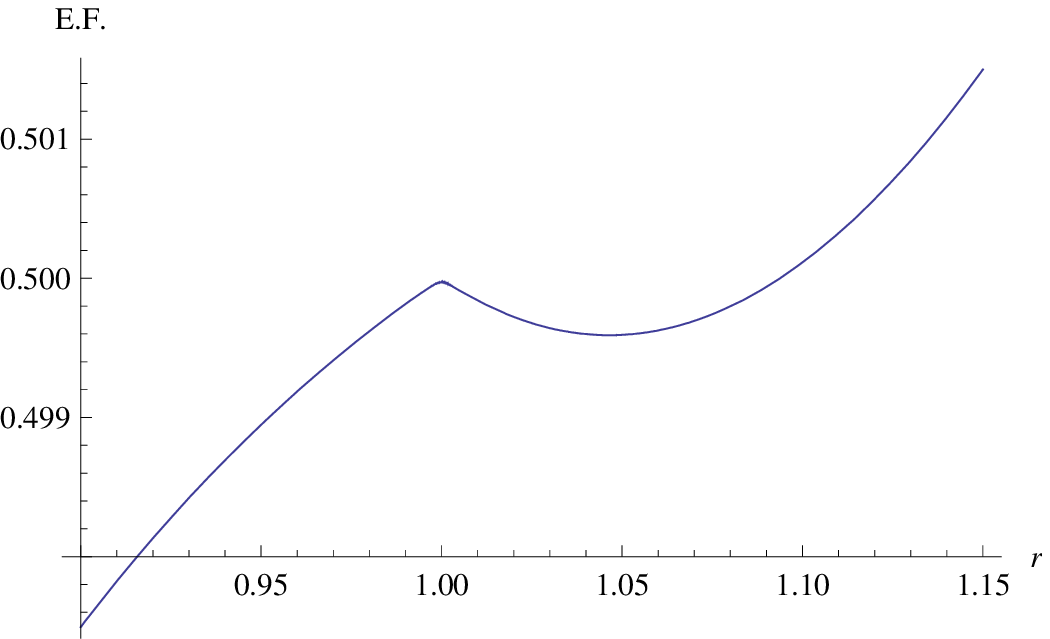}}
      }
      \label{fig:3}
     \end{minipage}%
     \hspace{3cm}
     \begin{minipage}[h]{8cm}
      \subfigure[]
      {
    \raisebox{ 0.45cm }{ 
      \resizebox{9cm}{!}{\includegraphics{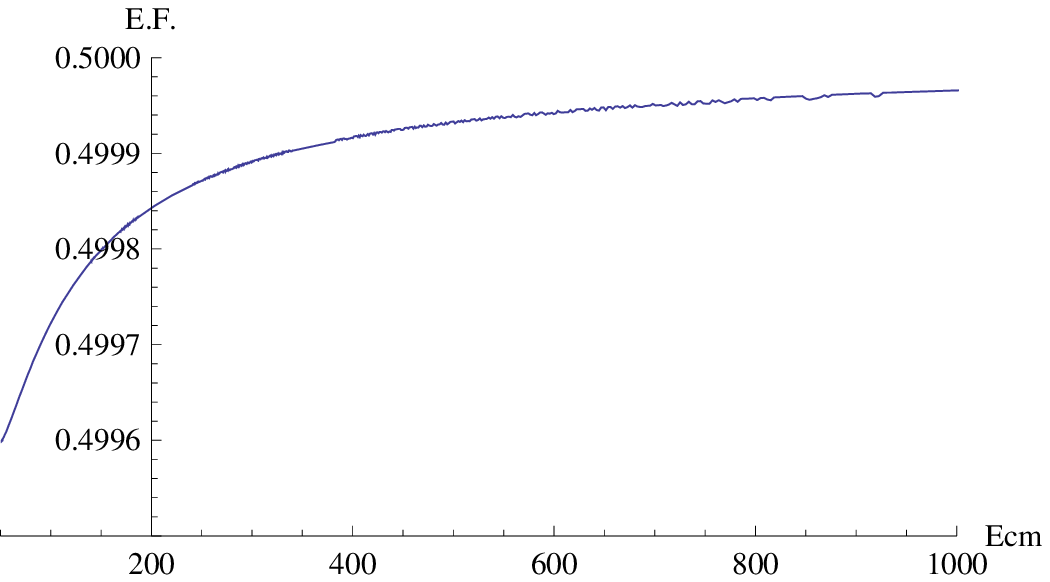}} }
      }
      \label{fig:4}
     \end{minipage}
\caption[]{Fig 3a shows the variation of escape fraction with
radius. Highly energetic collisions take place around $r=1$. 
Here $\epsilon=a-1=10^{-6}$. The collision is between
ingoing and outgoing particles with angular momenta $L_1=0.4,L_2=1.4$. The
escape fraction is more or less constant and takes a value
$E.F.\approx0.5$, unlike the extremal Kerr black hole case where it
decreases as one approaches the horizon $r=1$. Fig 3b depicts
the slight increase in the escape fraction with the center of mass
energy of collision, unlike the Kerr black hole case where
there is a sharp fall in the escape fraction with the center of
mass energy of collision.}
\end{figure*}

We assume that the massless particles produced in the collisions 
travel in the equatorial plane and their distribution 
is isotropic in the center of mass energy frame.
Particles would escape to infinity if \eq{escape} are satisfied. 
These conditions can be translated into the escape cones in the
center of mass frame such that if the direction of motion (in other 
words three-velocity) 
lies in the cone, particle escapes to infinity.
Directions that define edges of the escape cones would correspond
to the extreme angular momentum values in \eq{escape} given the 
radial location of collision 
and whether particle is ingoing or outgoing.
Given the extreme momentum values one can write down the 
velocity components of the massless
particle in the Boyer-Lindquist coordinate system. Then one can 
make a transition to the center of mass frame
and write down the three-velocities as
\begin{equation}
U_{3}^{i}=U^{\hat{i}}=\Lambda^{\hat{i}}_{~\tilde{\mu}}e^{~\tilde{\mu}}_{\nu}U^{\nu}.
\end{equation}
Angle between two extreme three-velocities will be the escape angle. 
When escape angle is divided by $2\pi$
we get the escape fraction,
\begin{equation}
EF(r,a)=\frac{\Theta(r-r_{max}(a))A_{1}(r,a)+\Theta(r_{max}(a)-r)A_{2}(r,a)}{2\pi}
\nonumber
\end{equation}
Here $\Theta$ is the step function.
The first term makes a contribution if collision takes place 
outside $r_{max}(a)$, whereas the second term contributes if collision
takes place inside $r_{max}(a)$. Since we are dealing with the case 
where $a\rightarrow 1$ in that case the high energy collisions will 
take place around $r=1$. In this situation $r_{max}(a)\simeq 4$ and 
the second term is relevant here but not the first one. There are two 
escape cones and thus there are two terms in $A_2$.
\begin{widetext}
\begin{eqnarray}
\nonumber
A_{1}(r,a)=
\left| \left(\arccos{\left[\frac{U_{3}(r,a,L=L_{1}(r),u=1).U_{3}(r,a,L=L_{2}(r),u=1)}{\mid U_{3}(r,a,L=L_{1}(r),u=1)\mid\mid U_{3}(r,a,L=L_{2}(r),u=1)\mid}\right]}\right)\right| \\
\nonumber
+\left| \left(\arccos{\left[\frac{U_{3}(r,a,L=a,u=-1).U_{3}(r,a,L=L_{1}(0),u=-1)}{\mid U_{3}(r,a,L=a,u=-1)\mid\mid U_{3}(r,a,L=L_{0},u=-1)\mid}\right]}\right)\right| \\
\nonumber
+\left| \left(\arccos{\left[\frac{U_{3}(r,a,L=L_{2}(r),u=-1).U_{3}(r,a,L=L_{2max},u=-1)}{\mid U_{3}(r,a,L=L_{2}(r),u=-1)\mid\mid U_{3}(r,a,L=L_{2max},u=-1)\mid}\right]}\right)\right|
\nonumber
\end{eqnarray}
\begin{eqnarray}
\nonumber
A_{2}(r,a)=
\left| \left(\arccos{\left[\frac{U_{3}(r,a,L=L_{1}(r),u=1).U_{3}(r,a,L=L_{2max},u=1)}{\mid U_{3}(r,a,L=L_{1}(r),u=1)\mid\mid U_{3}(r,a,L=L_{2max},u=1)\mid}\right]}\right)\right| \\
\nonumber
+\left| \left(\arccos{\left[\frac{U_{3}(r,a,L=L_{1}(r),u=-1).U_{3}(r,a,L=a,u=-1)}{\mid U_{3}(r,a,L=L_{1}(r),u=-1)\mid\mid U_{3}(r,a,L=a,u=-1)\mid}\right]}\right)\right| \\
\nonumber 
\end{eqnarray}
\end{widetext}

We numerically calculate the escape fraction for high energy 
collisions that take place around $r=1$. The spin parameter is 
slightly larger than unity $\epsilon=a-1\rightarrow 10^{-6}$. 
Collision is between 
ingoing and outgoing particles with angular momenta $L_1=0.4,L_2=1.4$.
The escape fraction as a function of radial coordinate
is plotted in Fig 3a. Escape fraction in this region around radial 
coordinate $r=1$ takes a value which is approximately half.
Variation of the escape fraction with the center of mass energy 
of collision is depicted in Fig 3b.
Escape fraction increases slowly with the center of mass energy. 
This behavior is in contrast with the 
extremal Kerr black hole case where there is a sharp decrement 
in the escape fraction with center of mass
energy for the collision as we approach the horizon $r=1$ \cite{BSW2}.
Thus the flux of the particles produced in the high energy collisions 
around the 
Kerr naked singularity will be large as approximately half of the 
particles manage to escape to infinity
while remaining half of the particles eventually hit the naked 
singularity. The high energy collisions, unlike 
the black hole case, would be observable and in principle would 
shed light on the nature of basic interactions 
at very large energies.

\section{Conclusions and Discussion}
Collisions with large center of mass energy can take place 
around extremal Kerr black hole. 
But since they occur at the location close to the event horizon, so 
most of the collision products formed 
are absorbed by the black hole. Thus these collisions, although 
interesting from the theoretical perspective, 
are not observable at all. Ultrahigh energy collisions can also 
take place around Kerr naked singularity 
provided the spin parameter transcends the unity by extremely 
small amount. However, the collisions take place at a location
away from the singularity. Thus one might expect that the significant 
fraction of collision products would manage 
to escape to infinity. In this paper we calculated the escape 
fraction for the collision products in the Kerr naked singularity
case and confirmed that it is indeed a finite and large enough 
number, unlike the black hole
case. The escape fraction for high energy collisions
is approximately half in this case. This implies that half of the collision
products manage to escape to infinity. Thus high energy
collisions around naked singularity are observable, in principle, 
and could help us understand the laws of basic interesting 
physics at ultrahigh energies.

Another important issue as far as the observability of the 
collisions is concerned is the time
required for the collision to occur. We had shown earlier 
in the context of the Reissner-Nordstr\"{o}m spacetime that the 
timescale associated with the Planck scale collision in a reasonable 
astrophysical setting is significantly smaller than the Hubble time
in the naked singularity case,
while that for black hole is many orders of magnitude larger than 
the age of the universe \cite{Patil2}. Similar results 
also hold good in the context of Kerr geometry as well. We shall 
present these findings elsewhere \cite{Patil4}.

\end{document}